\documentclass[10pt,twocolumn,prl]{revtex4}
\usepackage{graphicx}
\usepackage{amssymb}
\usepackage{amsmath}
\usepackage{bm}
\usepackage{color}

\RequirePackage{pdf14}

\begin{document}

\title{Walk-off Controlled Self-Starting Frequency Combs in $\bm{\chi^{(2)}}$ Optical Microresonators}

\author{S.\ Smirnov$^1$, B.\ Sturman$^2$, E.\ Podivilov$^{1,2}$, and I.\ Breunig$^{3,4}$}
\affiliation{$^1$Novosibirsk State University, 630090, Novosibirsk, Russia \\
$^2$Institute of Automation and Electrometry, Russian Academy of Sciences,
630090 Novosibirsk, Russia \\
$^3$University of Freiburg, Department of Microsystems Engineering - IMTEK, Georges-K\"{o}hler-Allee 102, 79110 Freiburg, Germany \\
$^4$Fraunhofer Institute for Physical Measurement Techniques, 79110 Freiburg, Germany
}

\begin{abstract}

Investigations of frequency combs in $\chi^{(3)}$ optical microresonators are burgeoning nowadays.
Changeover to $\chi^{(2)}$ resonators promises further advances and brings new challenges. Here,
the comb generation entails not only coupled first and second harmonics (FHs and SHs)
and two dispersion coefficients, but also a substantial difference in the group velocities
-- the spatial walk-off. We predict walk-off controlled highly stable comb generation,
drastically different from that known in the $\chi^{(3)}$ case. This includes the general notion of
antiperiodic state, formation of coherent antiperiodic steady states (solitons), where the
FH and SH envelopes move with a common velocity without shape changes, characterization of
the family of antiperiodic steady states, and the dependence of comb spectra on the pump power
and the group velocity difference. \\

\end{abstract}

\date{\today}

\maketitle

Frequency combs~\cite{Hansch02,Book05} consisting of equidistant optical lines
are indispensable for metrology, spectroscopy, and other applications~\cite{Standards,Spectroscopy1,Spectroscopy2}.
During the last decade, microresonator comb concept becomes increasingly important. Generation of
high-quality frequency combs in $\chi^{(3)}$ microresonators, see~\cite{KippNature07,Octave11,KippScience11,Herr12,KippNP14,Vahala15,Vahala18,KippScience18}
and references therein, is one of the most spectacular recent achievements in nonlinear optics.
These combs correspond to continuous-wave single-frequency pumped coherent spatial structures circulating
along the resonator rim with a constant speed.  Typically, these structures are dissipative
solitons balancing not only dispersion broadening and nonlinearity, but also external pumping
and internal losses~\cite{KippNP14,KippScience18}. They can be substantially different from solitons
in conservative systems. The outstanding comb properties are due to high $Q$-factors and small
volumes of the resonator modes.

\begin{figure}[h]
\centering
\includegraphics[width=8.2cm]{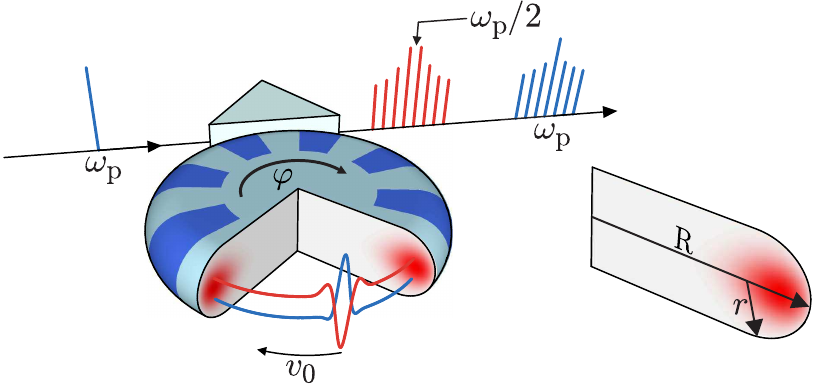}
\caption{Schematic of $\chi^{(2)}$ comb generation. Continuous-wave pump at the
frequency $\omega_p$ generates coupled SH and FH combs in a microresonator owing to
cascaded second-order nonlinear processes. These combs correspond to SH and FH solitons
moving along the resonator rim with a common velocity $v_0$. Red spots show localization
of the resonator modes; $\varphi$ is the azimuth angle. Radial poling and rim shaping
via control of the ratio of major ($R$) and minor (r) radii allow for quasi-phase matching
and mode management, respectively. }\label{Geometry}
\end{figure}

Transfer of the comb concept to $\chi^{(2)}$ resonators represents a big challenge and offers
new opportunities. Here, there are two light envelopes -- the first and second harmonics
(FH and SH) -- instead of one and, therefore, two dispersion coefficients.  Also there
is a substantial group velocity difference leading to the spatial walk-off between
FH and SH. Furthermore, phase matching has to be ensured for the second-order nonlinear processes.
Realization of $\chi^{(2)}$ combs promises lowering the pump power, entering new spectral ranges,
and new operation regimes. In particular, the presence of two subcombs in the FH and SH spectral
ranges, see Fig.~\ref{Geometry}, is a new positive feature. 

First attempts were undertaken to explore soliton regimes relevant to the frequency combs in
$\chi^{(2)}$ resonators~\cite{Wabnitz18,Skryabin19,Skryabin19A}. They concern with quadratic
nonlinear processes running at the spectral point of equal FH and SH group velocities. The found
dissipative solitons show a close relation to the conservative solitons reviewed in~\cite{SkryabinReview}.
Walk-off controlled soliton solutions at nonzero pump and zero losses were found~\cite{We19}; their 
stability is an open issue. Also, numerical results on analysis of comb regimes regardless of solitons 
are known~\cite{Att1,Att2,Att3,Att4}. The overall physical pattern of $\chi^{(2)}$ combs remains
obscure.

In this Letter, we report on walk-off controlled comb solutions for $\chi^{(2)}$ resonators
incorporating pump and losses and applicable to broad spectral ranges of the pump frequency.
The found solutions belong to the class of antiperiodic states, including solitons, which
are topologically different from conventional periodic states. Also they are totally different
from solutions of~\cite{We19}. The comb generation is stable against small and large perturbations, 
it can start from noise.

High-$Q$ microresonators possess discrete frequency spectra~\cite{Vahala03,IlchenkoReviewI,Review16}.
Optical modes of a resonator with major radius $R$ can be viewed as quasi-plane waves propagating along
the resonator rim and characterized by the azimuth angle $\varphi$~\cite{Vahala03,IlchenkoReviewI,Review16},
see Fig.~\ref{Geometry}. The modal functions are $2\pi$-periodic. Following all known theoretical comb
studies, we restrict ourselves to a single transverse mode type; this implies reducing transverse mode
number via rim shaping. Each mode is characterized by the azimuth number $m$ or by the wavenumber
$k_m = m/R$. For typical $\chi^{(2)}$ resonators, $m \sim 10^4$, and $k_m$ form a quasi-continuous set.
The modal frequencies are $\omega_m = 2\pi c/\lambda_m = k_mc/n(\lambda_m)$, where $c$ is speed of light,
$\lambda_m$ the vacuum wavelength, and $n(\lambda)$ the effective refractive index slowly varying with
$\lambda$. It is close to the bulk index $n_b(\lambda)$, but includes corrections relevant to the
geometric dispersion and vectorial coupling~\cite{Gorodetsky1,Gorodetsky2,ModalStructure,Vectorial}.

For quadratic nonlinearity, the phase-matching (PM) conditions $\omega_{k_2} = \omega_{k_1} + \omega_{k'_1}$,
$k_2 = k_1 + k'_1$ with discrete wavenumbers have to be fulfilled. At $k_1 = k'_1$ they give the SH
generation conditions. The latter are fulfilled in exceptional cases~\cite{NaturalPM}. However, equivalent
quasi-PM conditions can by ensured via the radial poling, see also Fig.~\ref{Geometry} and Supplemental
Materials~1 ({\color{red} SM1}), practically for any spectral range~\cite{RadialPoling1,RadialPoling2}.
This admits the presence of small frequency differences, such that $|\omega_{2k_1} - 2\omega_{k_1}| \ll c/nR$.
Continuous fine PM tuning means are also available~\cite{Review16,Tuning1,Tuning2}.

Let the PM conditions $\omega_{k^0_2} = 2\omega_{k^0_1}$, $k^0_2 = 2k^0_1$ be fulfilled and the pump
frequency $\omega_p$ be very close to $\omega_{k^0_2}$ (SH pumping). This means that the azimuth
number $m^0_2$ is even and $m^0_1 = m^0_2/2$ is an integer, see also Fig.~2a. The true light electric
field can be represented as
\begin{equation}\label{envelopes}
S\exp[{\rm i}(m^0_2\varphi -\omega_pt)] + F\exp[{\rm i}(m^0_2\varphi -\omega_pt)/2] + c.c.,
\end{equation}
where $F(\varphi,t)$ and $S(\varphi,t)$ are complex FH and SH envelopes, both $2\pi$-periodic
in $\varphi$. These envelopes obey a generic set of nonlinear equations~\cite{SkryabinReview,Skryabin19,We19}:
\begin{eqnarray}\label{Initial}
\hspace*{-2.5mm}&\bigg[& \hspace*{-2mm} i\bigg(\hspace*{-0.5mm}\frac{\partial}{\partial t} +
\frac{v_1}{R}\frac{\partial}{\partial \varphi} +\gamma_1 \hspace*{-1mm}\bigg) \hspace*{-0.5mm}
+ \hspace*{-0.5mm} \frac{v'_1}{2R^2}\frac{\partial^2}{\partial \varphi^2}
- \Delta_1 \bigg]F = 2\mu SF^* \\
\hspace*{-2.5mm}&\bigg[& \hspace*{-2mm} i\bigg(\hspace*{-0.5mm}\frac{\partial}{\partial t} +
\frac{v_2}{R}\frac{\partial}{\partial \varphi} + \gamma_2 \hspace*{-1mm}\bigg) \hspace*{-0.5mm}
+ \hspace*{-0.5mm} \frac{v'_2}{2R^2}\frac{\partial^2}{\partial \varphi^2}
- \Delta_2 \bigg]S = \mu F^2 + ih . \nonumber
\end{eqnarray}
Here $v_{1,2}$ and $v'_{1,2}$ are the group velocities and the dispersions (discrete equivalents
of $d\omega/dk$ and $d^2\omega/dk^2$) at $k^0_{1,2}$, $\gamma_{1,2}$ are the modal decay constants,
$\Delta_{1,2}$ are the frequency detunings accounting for a slightly imperfect PM
and a small difference $\omega_p - \omega_{k^0_2}$, $\mu$ is the coupling constant proportional to the second-order nonlinear coefficient, and $h$ is
the pump strength. These parameters are real and experimentally controlled; the ratios $v_{1,2}/2\pi R$
($\gtrsim 10$~GHz) are known as FH and SH free spectral ranges.

\begin{figure}[h]
\centering
\includegraphics[width=8.2cm]{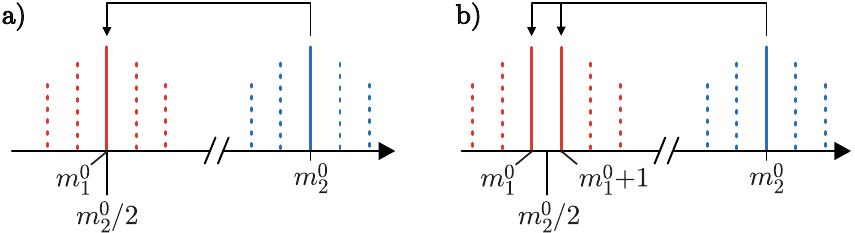}
\caption{Excitation of periodic (a) and antiperiodic (b) states at SH-pumping of even
and odd modes. The FH carrier frequency $m^0_2/2$ is integer in a) and semi-integer in
b). Side harmonics arise automatically in b) above the threshold.}\label{Antiperiodic}
\end{figure}

Now let us pump a SH mode with an odd azimuth number $m^0_2$, see also Fig.~2b. The PM conditions link
this mode to two FH modes possessing even and odd numbers: $\omega_{m^0_2} = \omega_{m^0_1} + \omega_{m^0_1
+ 1}$, $m^0_2 = 2m^0_1 + 1$. Is set~(\ref{Initial}) for $F$ and $S$, as defined by Eqs.~(\ref{envelopes}),
valid in this case? The answer is {\em yes}, but the FH envelope becomes antiperiodic, $F(\varphi) =
-F(\varphi + 2\pi)$. This follows from $2\pi$-periodicity of the true light field and the
presence of the antiperiodic factor $\exp({\rm i}m^0_2\varphi/2)$ in Eq.~(\ref{envelopes}).
The $2\pi$-periodic squared modulus $|F|^2$ still represents the true FH field intensity, while the
Fourier expansion of $F(\varphi)$ consists of semi-integer harmonics $F_{j_1}$ with $j_1 = m_1 - m^0_1
- 1/2 = \pm 1/2,\pm 3/2, \ldots$. The SH amplitude $S(\varphi)$ remains $2\pi$-periodic; it includes
only integer harmonics $S_{j_2}$ with $j_2 = m_2 - m^0_2 = 0, \pm 1, \ldots$. Attempts to employ
$2\pi$-periodic FH envelopes break the generic structure of Eqs.~(\ref{Initial}). Further details
can be found in {\color{red} SM2} and~\cite{We19}.

The antiperiodic solutions of Eqs.~(\ref{Initial}) are topologically different from the periodic ones.
They form a separate class of nonlinear states, which is specific for SH pumping and favorable
for comb generation. The differences between the periodic and antiperiodic states are crucial: \\
-- In the periodic case, there are spatially uniform solutions $\bar{F}(\varphi), \bar{S}(\varphi)
= const$. In the antiperiodic case, such solutions are impossible. Nevertheless, $\bar{F}$ and $\bar{S}$ represent here asymptotic almost $\varphi$-independent values of $F(\varphi)$ and $S(\varphi)$, see below.  \\
-- Harmonics $F_{1/2}$ and $F_{-1/2}$ not only influence $S_{0}$, but force harmonics $S_{\pm 1}$ 
enriching the SH spectrum. The latter cause new nonlinear processes, so that a broad FH spectrum
appears above a single oscillation threshold.

Nonlinear set~(\ref{Initial}), including many variable parameters, is generally very
complicated. In particular, it is more complicated compared to the Lugiato-Lefever equation~\cite{LL-equation,LL-SouthAfrica}
relevant to $\chi^{(3)}$ comb modeling~\cite{KippNP14,KippScience18}. We restrict ourselves to the case
of zero detunings $\Delta_{1,2} = 0$. This means that the PM conditions are fulfilled exactly.
Also, we set for simplicity $\gamma_{1,2} = \gamma$. Four dimensionless parameters control then the
nonlinear behavior. These are the normalized pump strength $\eta = 2\mu h/\gamma^2$, the walk-off parameter
$\alpha = v_{12}/\gamma R$ with $v_{12} = v_1 - v_2$, and two dispersion parameters $\beta_{1,2} =
v'_{1,2}/2\gamma R^2$.

\begin{figure}[h]
\centering
\includegraphics[width=8.6cm,height=3.2cm]{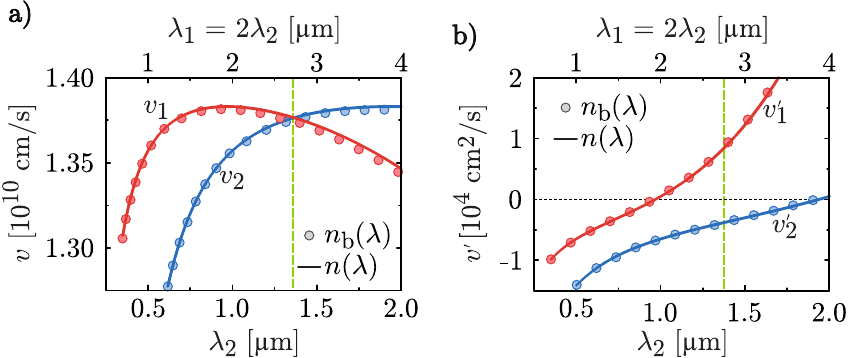} \vspace*{-3mm}
\caption{Wavelength dependences of the group velocities $v_{1,2}$ (a) and the dispersions
$v'_{1,2}$ (b) for lithium niobate based resonator with $R = 1.5$~mm and $R/r = 3$.
The solid and dotted lines refer to the bulk refractive index $n_b$ and the effective
index $n$ incorporating the effects of geometric dispersion. The verical lines $\lambda = \lambda_2^0 = \lambda_1^0/2 \simeq 1.36\,\mu$m correspond to $v_1 = v_2$.
}\label{Dispersions}
\end{figure}

Importance of different parameters and the physical pattern can be clarified taking into
account representative dependences $v_{1,2}(\lambda)$ and $v'_{1,2}(\lambda)$, see Fig.~\ref{Dispersions}.
For mm-sized resonators, the effects of geometric dispersion are weak. The group velocity difference
$v_{12} = v_1 - v_2$ ranges from huge values ($\sim 10^8$~cm/s) to zero at $\lambda^0_2 \simeq 1.36\,\mu$m.
Setting $R = 1.5$~mm, $\gamma = 10^7$~s$^{-1}$ ($Q \approx 10^8$), we get for the pump wavelength $\lambda_2 \simeq 1\,\mu$m: $\alpha \approx 2 \times 10^2$, $\beta_1 \approx 3 \times 10^{-3}$, and $\beta_2 \approx -3 \times 10^{-2}$. This means that the walk-off effects dominate over the dispersion ones. When moving 
to the point of equal group velocities $\lambda^0_2$, the coefficient $\alpha$ tends to zero, while
$\beta_1$ and $\beta_2$ remain opposite in sign with $|\beta_{1,2}| \approx 10^{-2}$. Thus, the walk-off 
effects can be controlled by the choice of the pump wavelength $\lambda_2$; they are small in the close vicinity of $\lambda^0_2$. The threshold value of $\eta$ for generation of the antiperiodic states is $\eta_{\rm th} = (1 + \beta^2_1/16)^{1/2} \simeq 1$, see also {\color{red} SM2}. For the periodic states, it is $\eta_{\rm th} = 1$. Thus, $\eta$ is expressible by the ratio of the pump power $\mathcal{P}$ to its 
threshold value: $\eta^2 = \mathcal{P}/\mathcal{P}_{\rm th}$.

Above the threshold, we are eager for steady states $F(\varphi - v_0t/R)$, $S(\varphi - v_0t/R)$
moving with a common velocity $v_0$ without shape changes. Only such states provide FH and SH
frequency combs; the Fourier components $F_{j_1}$ and $S_{j_2}$ represent here the FH and SH frequency
spectra with the common frequency spacings $\delta \omega = v_0/R$. Periodic steady states with 
such frequency spacing are not expected for $\Delta_{1,2} = 0$: The spatially uniform solution 
$\bar{F},\bar{S}$ with $|\bar{F}|^2 \propto \eta - \eta_{\rm th}$ is known
to be stable here against spatially uniform and quasi-uniform perturbations. The {\em antiperiodic}
steady states provide potentially the best possibility for comb generation. The presence of such states 
and their stability against temporal perturbations are not granted. Also, velocity $v_0$ is unknown, it must be determined simultaneously with the shape of the steady states. Since the parametrically generated satellites $F_{\pm 1/2}$ propagate at the threshold with velocity $v_1$ and force an SH pattern propagating with the same velocity, we expect that $v_0 \simeq v_1$ near the threshold. The velocity difference $v_{01} = v_0 - v_1$ is an important parameter; the ratio $v_{01}/2\pi R$ characterizes fine nonlinear tuning of the comb frequency spacing.

We simulated numerically the Fourier transform of Eqs.~(\ref{Initial}) at $\Delta_{1,2}
= 0$ in the coordinate frame moving with velocity $v_1$ using the fourth-order
Runge-Kutta method. The total number of harmonics taken into account ranged from $64$ to
$512$. The harmonics $F_{j_1}(t)$ and $S_{j_2}(t)$ were found within a large
range of $j_{1,2}$ and within a sufficiently broad range of $\eta$, $\alpha$: $1 \leq \eta 
\leq 100$, $10^{-2} \leq \alpha \leq 10^2$. Accuracy of the calculations was controlled by changing the 
time step and the total number of harmonics. With the harmonics calculated, one can ensure 
establishment of the antiperiodic steady states and determine the velocity difference $v_{01}$, 
the comb spectra $|F_{j_1}|^2$ and $|S_{j_2}|^2$, and the spatial profiles $F(\varphi, t)$, 
$S(\varphi, t)$. 

The following quasi-adiabatic calculation procedure was used to cover the whole
$\eta,\alpha$ range: Parameter $\eta$ increased and decreased stepwise at certain
$\alpha$, and the previous values of $F_{j_1}$ and $S_{j_2}$ were used as new
initial conditions. Achievement of steady states was controlled in each step. It was
found that the result of temporal evolution does not depend on the initial conditions 
(including random noise). It is arrival at a unique coherent FH-SH pattern moving with 
a common velocity $v_0(\alpha,\eta)$. The sufficient restriction on the pump rise time $t_p$ is 
$t_p \gtrsim 1\,\mu$s; it is very soft. The found features mean that the comb generation
is self-starting. More details can be found in {\color{red} SM3.} Note that adiabatic changes 
of $\eta$ (of the pump power) are practical for experiment, whereas 
big adiabatic changes of $\alpha$ would be impractical.

Turning to the results, we start with dependences of the comb spectra on $\alpha$ and
$\eta$. For the FH and SH amplitudes, it is convenient to use the normalized quantities
$f = (\mu/\gamma) (2i)^{1/2}F$ and $s = (2\mu/\gamma) S$. With this normalization, the
spatially uniform steady states are given by $\bar{f} = \pm (\eta - 1)^{1/2}$ and $\bar{s}
= 1$. This sets a useful reference scale. While all harmonics are nonzero in steady
state, we restrict ourselves to the range $|f_{j_1}|^2,|s_{j_2}|^2 \geq 10^{-4}$ when
presenting the comb spectra. The corresponding numbers of significant FH and SH comb
lines we denote $N_1$ and $N_2$. 

\begin{figure}[h]
\centering
\includegraphics[width=9cm]{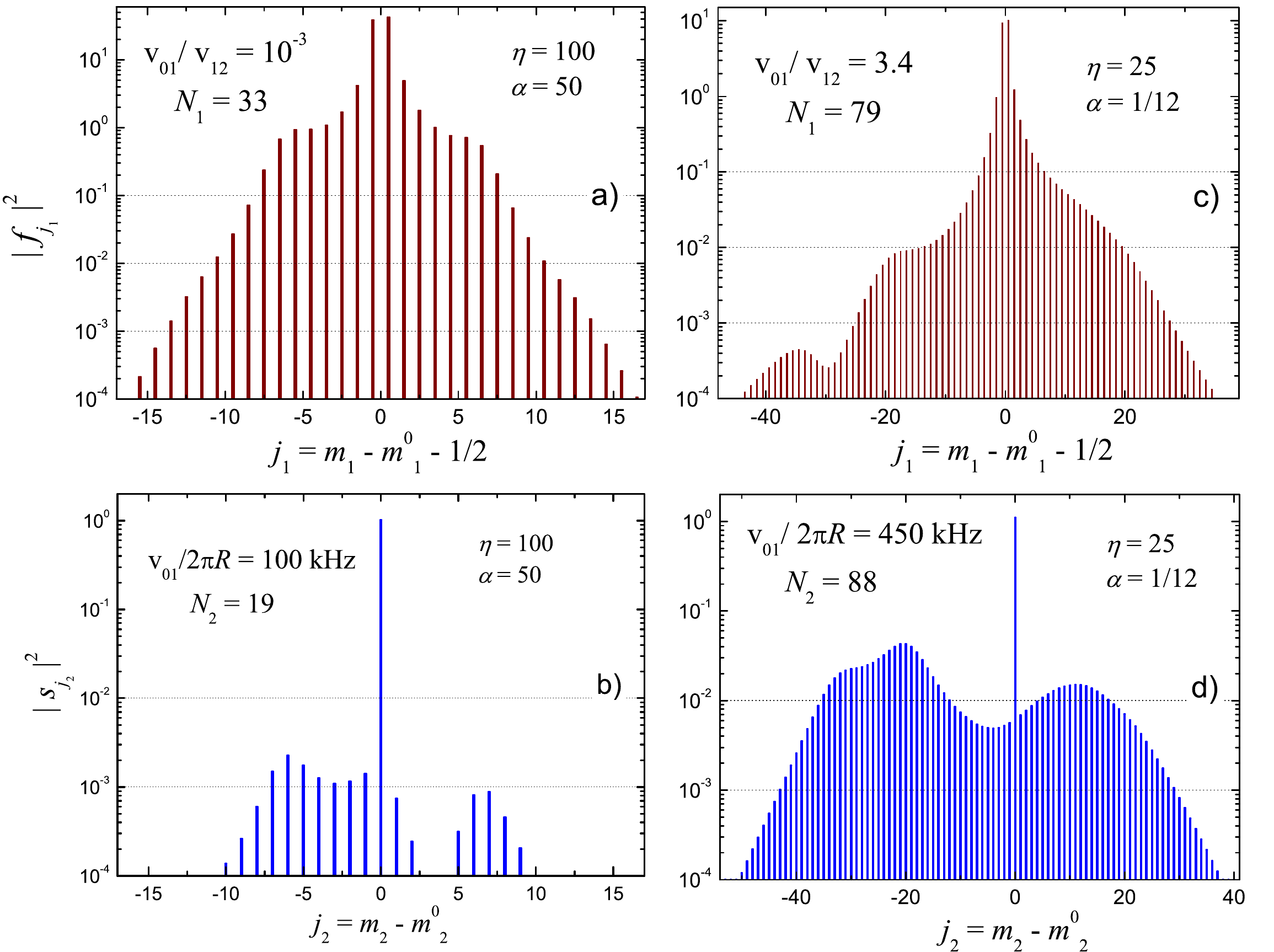}
\caption{Steady-state comb spectra $|f_{j_1}|^2$ and $|s_{j_2}|^2$ for two combinations
of $\eta$ and $\alpha$ and the dispersion parameters $\beta_1 = 0.02$ and $\beta_2 = -
0.01$. a),b): $\eta = 100$ and $\alpha = 50$; c),d): $\eta = 25$ and $\alpha = 1/12$.
Only the comb lines above $10^{-4}$ are shown. The total number of Fourier harmonics
taken into account is 128 for a),b) and 512 for c),d). The frequency spacing between
the lines is $\delta \omega = v_0/R$.}\label{CombSpectra}
\end{figure} 

Figure~\ref{CombSpectra} shows the normalized comb spectra for two combinations of $\eta$ 
and $\alpha$ and the dispersion parameters $\beta_1 = 0.02$, $\beta_2 = -0.01$.
For $\alpha = 50$ and $\eta = 100$, represnting large $\alpha$ and $\eta$, we have $N_1 = 33$, $N_2 = 19$, 
see Figs.~\ref{CombSpectra}a,b, a very small positive velocity ratio $v_{01}/v_{12} \approx 10^{-3}$, 
and $v_{01}/2\pi R \approx 0.1$~MHz. In the SH spectrum there is one dominating line, 
$|s_0|^2 \simeq 1$. For $\alpha = 1/12$ and $\eta = 25$, representing small walk-off parameters and modest pump strengths, we have well developed FH and SH spectra with $N_1 = 79$ and $N_2 = 88$, see Figs.~\ref{CombSpectra}c,d, corresponding to $v_{01}/v_{12} \approx 3.4$ and $v_{01}/2\pi R \approx 0.45$~MHz. Domimation of $|s_0|^2$ over the SH wings is much less pronounced. The left-right asymmetry of the spectra of Fig.~\ref{CombSpectra}c,d and their ripple structure are remarkable; these features are due to an interplay between the walk-off and dispersion effects. 

\begin{figure}[h]
	\centering
	\includegraphics[width=8.6cm]{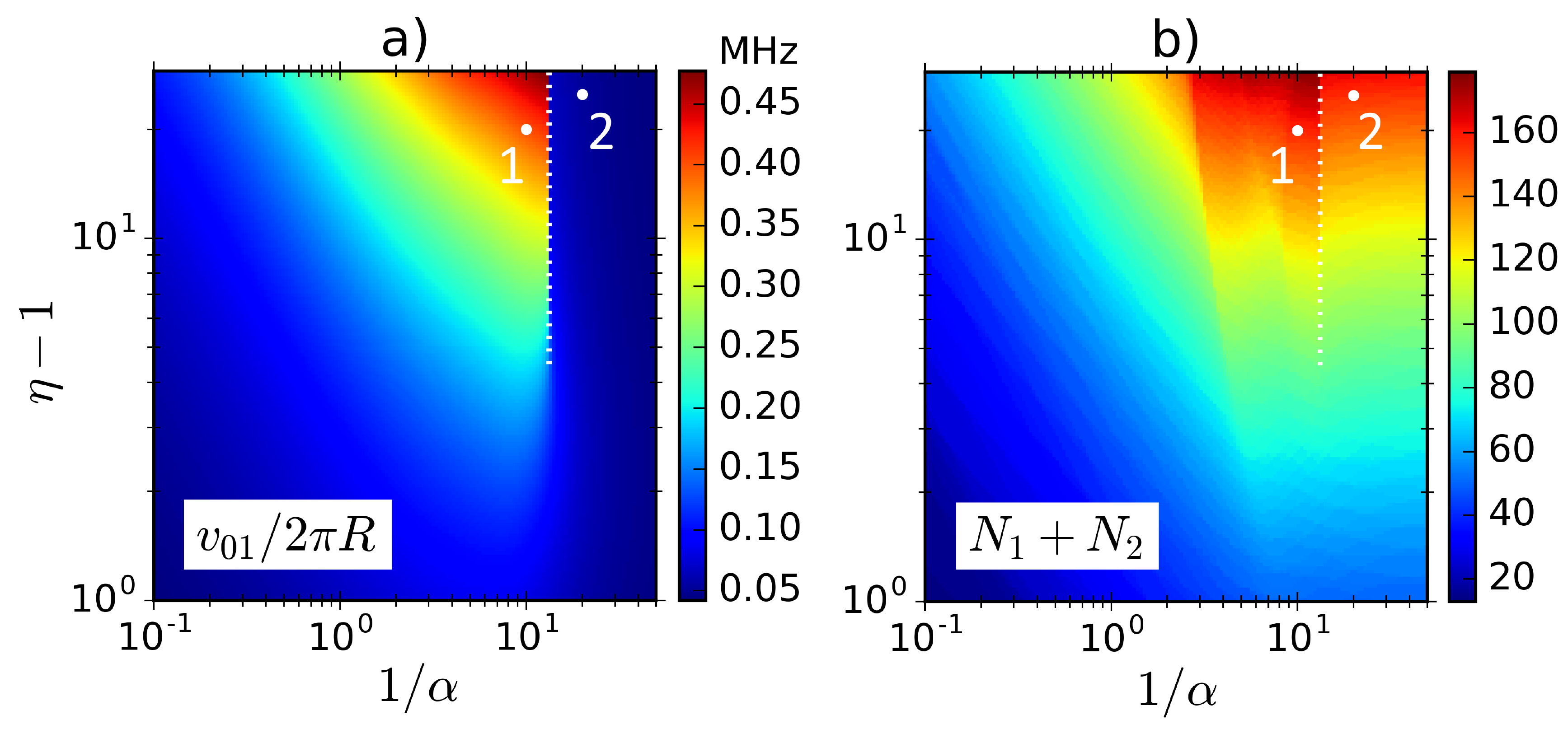}
	\caption{Maps of the tuning parameter $v_{01}/2\pi R$ (a) and of the total number of
		significant comb lines $N_1 + N_2$ (b) on the $\alpha^{-1},\eta$ plane for $\beta_1 = 0.02$ 
		and $\beta_2 = -0.01$. The vertical
		line of discontinuity starts at $1/\alpha_c \simeq 13$, $\eta_c \simeq 4.5$. Each map
		incorporates the data of $151 \times 151 = 22801$ calculation variants with 512
		harmonics taken into account. White dots 1 and 2 correspond to the points of the plane 
		$(10,20)$ and $(20,25)$ lying to the left and right of the discontinuity and also
		to profiles ${\bf 1}$ and ${\bf 2}$ in Fig.~6.}\label{Maps}
\end{figure}

Next, we consider the tuning parameter $v_{01}/2\pi R$ and the total number of comb
lines $N_1 + N_2$ as functions of $\alpha^{-1},\eta$ within the range $0.1 \leq \alpha^{-1} \leq 50$, $1 \leq \eta \leq 30$, see Fig.~\ref{Maps}. A remarkable
feature is here the presence of the vertical line of discontinuity $\alpha = \alpha_c
\simeq 1/13$, $\eta > \eta_c \simeq 4.5$.

Both mapped quantities grow with increasing $\eta$, but this growth is substantially
weaker for $\alpha < \alpha_c$. Taken as functions of $\lg(1/\alpha)$, they grow first
approximately linearly and then drop and stop growing. The drops in a) and b) are
relatively large and small. Furthermore, $v_{01}$ tends to zero for $\alpha \to 0$.
In essence, the discontinuity marks a sharp transition from the walk-off to dispersion
controlled comb regimes for $\eta > \eta_c$. For $\eta < \eta_c$, this transition occurs
continuously with increasing $\lg(1/\alpha)$. When moving up along the left side of the
discontinuity, the tuning parameter approaches the MHz range for $\eta \approx 30$,
and the values of $v_{01}/v_{12}$ and $N_1 + N_2 \simeq 2N_1$ approach $3.4$ and $180$,
respectively. Further increase of $\eta$ presents no numerical difficulties, but can lead to excessively 
large pump powers. Minor irregularities of the map b) are caused by discreteness of $N_{1,2}$ 
and by the ripple structure of the comb spectra: small variations of $\eta$ and $\alpha$ 
cause sometimes stepwise changes of these integers. 

\begin{figure}[b]
\centering
\includegraphics[width=8.5cm]{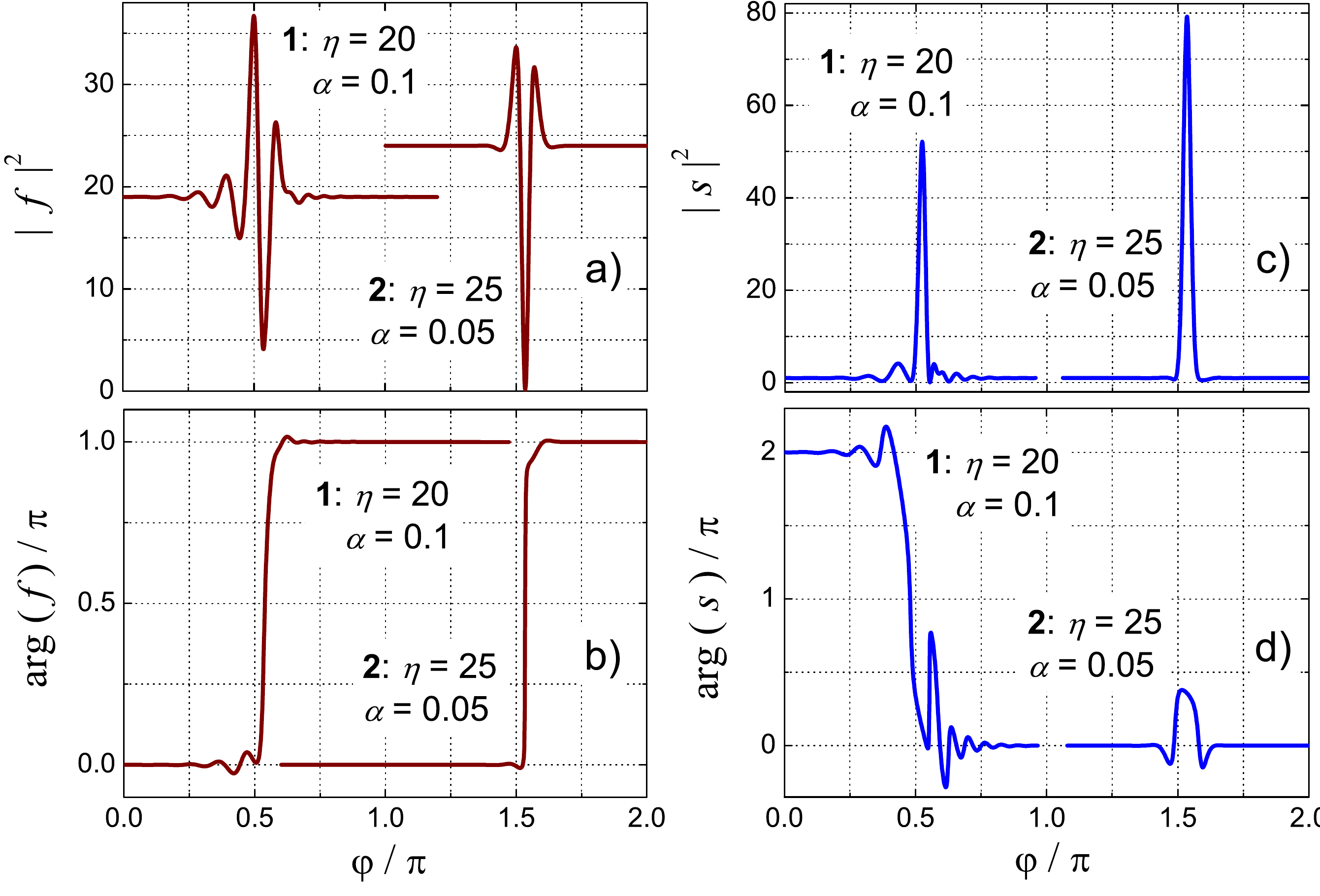}
\caption{Antiperiodic solitons for points ${\bf 1}$ and ${\bf 2}$ lying to the left and right of the discontinuity in Fig.~5. Sub-figures a) to d) show $|f(\varphi)|^2$, $\arg [f(\varphi)]$, $|s(\varphi)|^2$, 
and $\arg[s(\varphi)]$. The background values of $|f|^2$ and $|s|^2$ are $\eta - 1$, and $1$,
respectively. The $\pi$-steps of $\arg[f(\varphi)]$ in b) occur at the points of minimum
of $|f(\varphi)|^2$ in a). Changes of $\arg[s(\varphi)]$ occur near the maxima of $|s(\varphi)|^2$, 
they are qualitatively different for ${\bf 1}$ and ${\bf 2}$. 512 FH and SH harmonics
are taken into account.}\label{SpatialProfiles}
\end{figure}

Consider the spatial structure of our antiperiodic steady states.
Figure~\ref{SpatialProfiles} shows the FH and SH intensities and phases versus the azimuth angle $\varphi$ for points ${\bf 1}$ and ${\bf 2}$ on the 
$\alpha^{-1},\eta$ plane lying slightly to the left and right of the discontinuity in Fig.~5.
It is evident from a) and c) that the intensity changes are strong and well localized,
so that we are dealing with FH and SH solitons propagating with velocity $v_0$. Each
intensity distribution has a background; in accordance with the previous considerations,
the FH and SH intensity backgrounds are $\bar{f}^{\,2} = \eta - 1$ and $\bar{s}^{\, 2} =
1$. The intensity profiles ${\bf 1}$ are much more oscillatory as compared to  
profiles ${\bf 2}$. The phase $\arg[f(\varphi)]$ exhibits sharp $\pi$-steps in both cases 
ensuring the antiperiodic behavior. The behavior of $\arg[s(\varphi)]$ is, however, not uniform: 
the phase profiles ${\bf 1}$ and ${\bf 2}$ in Fig.~\ref{SpatialProfiles}d show $2\pi$-drop and 
a zero overall change. 

What happens with the solitons when changing $\eta,\alpha$? Decrease of $\alpha$ as
compared to $\alpha_c$ gives no effect, the profiles ${\bf 2}$ in Fig.~6 correspond
practically to the limit $\alpha \to 0$. Increasing $\alpha$ causes a weak bifurcation
of soliton ${\bf 1}$: the $2\pi$-drop of $\arg(s)$ changes to $0$. This bifurcation
is not accompanied by discontinuities of $v_{01}$ and $N_{1,2}$. 

Above we focused on the spectral range $\lambda_2 < \lambda_2^0 \simeq 1.36\,\mu$m, where
$v_{12}(\lambda_2) > 0$, see Fig.~\ref{Dispersions}. Generalization to the range $\lambda_2 >
\lambda_2^0$, where $v_{12} < 0$, presents no difficulties. It results in changing sign
of $v_{01}$. Modest variations of the dispersions $\beta_{1,2}$ and of the ratio
$\gamma_1/\gamma_2$ influence quantitative details, but not the physical pattern.

How precise and informative is our assertion about achievement of the steady states? To
clarify it, we introduce the discrepancy parameter
\begin{equation}\label{Norm}
\varepsilon(t,\tau) = \frac{\sum_j \left|A_j(t) - A_j(t - \tau)\right|^2}{\sum_j \left( |A_j(t)|^2
+ |A_j(t - \tau)|^2 \right)},
\end{equation}
where $A_j$ is one of the harmonics $F_{j_1}$ and $S_{j_2}$ in the frame moving with an
arbitrary velocity $v$, $t$ is the calculation time, and $\tau$ is a variable time
shift. As soon as harmonics $A_j(t)$ are known in the frame moving with velocity $v_1$,
they can be recalculated in the frame moving with velocity $v$ through multiplication by
$\exp[- ij_{1,2}(v - v_1) t/R]$. Obviously, $\varepsilon (t,\tau)$ turns to zero only when we
deal with the steady state and, simultaneously, $v = v_0$. The discrepancy parameter
calculated for modestly large evolution times, $\gamma t \gtrsim 10^3$, and minimized
over $v$ shows extremely small values ($\varepsilon = 10^{-14} - 10^{-15}$) caused by the 
numerical noise, see {\color{red} SM3} for details. For smaller $t$, i.e., during the
transient stage, it is larger by many orders of magnitude. Thus, we have a tool to
control proximity of the steady states and to determine precisely velocity $v_0$. 

For $\eta \gg 1$ and an abrupt (non-adiabatic) turning the pump on, the above scenario
of nonlinear evolution to unique steady states can be violated. In this case, generation
of single-soliton steady states from noise occurs probablistically, and complicated
multi-soliton structures become most probable.

Turning to discussion, we consider first the main distinctive features of this study: \\
-- The results found for $\chi^{(2)}$ resonators concern with new antiperiodic nonlinear
comb states that are topologically different from conventional periodic states. To
excite the antiperiodic states, it is sufficient to pump SH modes with odd azimuth
numbers. Neither $\chi^{(3)}$ nor FH pumped $\chi^{(2)}$ resonators possess such states.
The antiperiodic states are the most favorable for $\chi^{(2)}$ comb generation:
Formation of broad spectra of Fourier harmonics occurs automatically above
a single optical oscillation threshold, while spatially-uniform background states
are forbidden. \\
-- The necessary condition for the generation of equidistant $\chi^{(2)}$ frequency
combs, formation of FH and SH envelopes propagating with a common constant velocity without
shape changes, is fulfilled in broad
ranges of experimental parameters, such as pump wavelength and power. Moreover, the
comb states are self-starting -- the nonlinear evolution leads above the threshold
to a unique comb state under weak limitations on the pump rise time. \\
-- In contrast to the previous $\chi^{(2)}$ comb studies~\cite{Wabnitz18,Skryabin19,Skryabin19A},
we are not attached to the spectral point of equal FH and SH group velocities $\lambda^0_2$.
The spatial walk-off of the FH and SH envelopes, caused by generic group velocity difference,
controls the spectral features of the comb solutions together with the pump power.    \\
-- Broad comb spectra correspond to a vast family of spatially narrow antiperiodic
dissipative solitons. These solitons not only balance the dispersion broadening and nonlinear
narrowing, gain and losses, but also ensure a common velocity of FH and SH envelopes. 
To the best of our knowledge, this multiparametric soliton family has no analogues in the literature. \\

While the single-mode and perfect radial poling assumptions are realistic and common for the theoretical comb studies, they are not always fulfilled in experiment. Special efforts are necessary thus for the experimental realization of $\chi^{(2)}$ combs. As concern our assumption of zero FH and SH frequency detunings, it is made for simplicity. On the one hand, it is realizable in experiment. On the other hand, admission of nonzero detunings is expected to strongly complicate the comb regimes. Possibly, it will results in new important predictions for experiment.

In conclusion, the presented theoretical results form a broad and solid frame for
practical realization of frequency combs in $\chi^{(2)}$ microresonators. They have a
big potential for further extension by incorporating the effects of FH and SH frequency
detunings.

\vspace*{1mm}
{\bf Acknowledgements:} This work was financially supported by RFBR, grant number 20-02-00511, and by the Fraunhofer and Max Planck Cooperation Program COSPA

\end{document}


\title{Supplemental Materials to \\ Walk-off Controlled Self-Starting Frequency Combs in 
$\bm{\chi^{(2)}}$ Optical Microresonators}

\author{S.\ Smirnov$^1$, B.\ Sturman$^2$, E.\ Podivilov$^{1,2}$, and I.\ Breunig$^{3,4}$}
\affiliation{$^1$Novosibirsk State University, 630090, Novosibirsk, Russia \\
$^2$Institute of Automation and Electrometry, Russian Academy of Sciences,
630090 Novosibirsk, Russia \\
$^3$University of Freiburg, Department of Microsystems Engineering - IMTEK, Georges-K\"{o}hler-Allee 102, 79110 Freiburg, Germany \\
$^4$Fraunhofer Institute for Physical Measurement Techniques, 79110 Freiburg, Germany }

\date{\today}

\maketitle

\section{The effect of radial poling}

\begin{figure}[h]
\centering
\includegraphics[width=12cm]{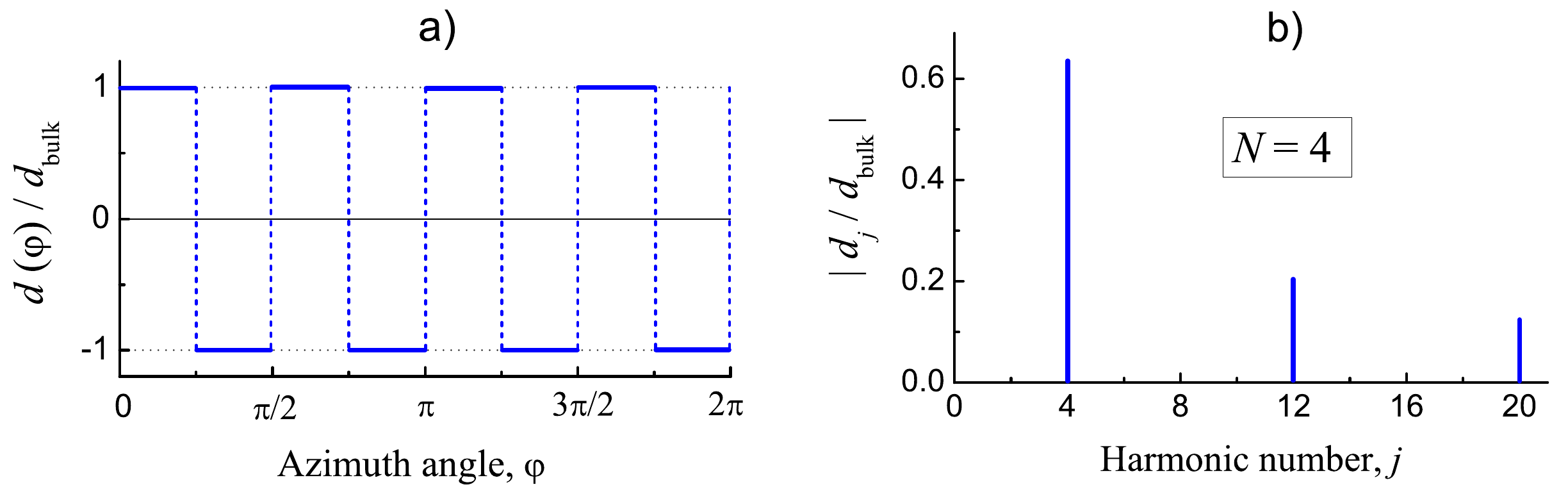}
\caption{a) Schematic of the azimuth dependence of the susceptibility coefficient $d(\varphi)$ 
for a perfect symmetric radial domain pattern with $N = 4$. b) First three harmonics of the 
corresponding Fourier spectrum with $j = N$, $3N$, and $5N$.} \label{Geo}
\end{figure}

The nonlinear response of ferroelectric $\chi^{(2)}$ materials, like LiNbO$_3$ or
LiTaO$_3$, is determined by the independent real $d_{333}$ and $d_{311}$ components of
the third-rank quadratic susceptibility tensor $\hat{d}$~\cite{Boyd}. These components change 
sign under inversion of the direction of the spontaneous polarization. In the case of perfect
radial poling, shown schematically in Fig.~1 of the main text, any of these components
(let it be $d$ with the bulk value $d_{\rm bulk}$) alternates periodically in a stepwise
manner with the azimuth angle $\varphi$, as illustrated in Fig.~\ref{Geo}a. At the same time, 
the linear susceptibility tensor and the linear optical properties stay unchanged. If $N$ 
is the number of alternation periods, the function $d(\varphi)$ is $2\pi/N$-periodic. 
It can be expanded in the Fourier series
\begin{equation}\label{Fourier}
d = \sum\limits_{j} d_j \exp({\rm i}j\varphi) \;,
\end{equation}
with $j = 0, \pm N, \pm 2N, \ldots$ In the case of $\pm$-symmetric radial domain structure,
which is the most suitable for quasi-phase matching, only the Fourier harmomics with odd ratios 
$j/N = \pm 1, \pm 3, \ldots$ are nonzero~\cite{RadialPoling1,IngoReview}. For these harmonics we have 
$|d_j| = 2d_{\rm bulk}N/\pi|j|$. The reduction factor $|d_j|/d_{\rm bulk}$ decreases with increasing 
$|j|/N$, but remains comparable with $1$ for $|j| = N$, see also Fig.~\ref{Geo}b.

Employment of the first Fourier harmonics of $d(\varphi)$ for quasi-phase matching
corresponds to the SH phase-matching conditions $2\omega_{m_1} = \omega_{2m_1 \pm N}$.
The sign "plus" is relevant to the most typical case of decreasing wavelength dependence
of the refraction index $n(\lambda)$~\cite{Nikogosyan}. We see that the SH azimuth number 
$m_2 = 2m_1 \pm N$ is even for even alternation number $N$ and odd for odd number $N$. In this case, 
SH pumping leads to the excitation of periodic states. In the case of odd combination $m_2
\mp N$, the SH quasi-phase matching condition cannot be fulfilled. However, we can
fulfil the quasi-phase matching condition $\omega_{m_1} + \omega_{m_1 + 1} =
\omega_{2m_1 \pm N}$ leading to the excitation of the antiperiodic states. The values of
$N$ typical for the radial poling are of the order of $10^2$~\cite{RadialPoling1,RadialPoling2}.

Thus, employment of perfect radial poling leads merely to the replacement of the SH
azimuth number $m_2$ by $m_2 \mp N$ and to the replacement of the bulk nonlinear
coefficient $d_{\rm bulk}$ by a slightly reduced coefficient $d_{\rm eff} = 2d_{\rm
bulk}/\pi$. Generally speaking, one can employ the higher Fourier harmonics for
quasi-phase matching. At the same number $N$, they correspond to substantially different
wavelength ranges and also smaller nonlinear coupling coefficients. It is worthy of
mentioning that the quasi-phase matching can be combined with continuous fine frequency
tuning, such as the temperature tuning or geometric tuning. This enables one to
accomplish the quasi-phase matching practically in any desirable spectral range.

Any disturbance of the $2\pi/N$ periodicity of the radial domain structure leads to
decrease of the primary Fourier harmonics $d_N$, $d_{3N}$, \ldots and also to additional
Fourier harmonics $d_j$ with $j \neq N$, $3N$, \ldots~\cite{RadialPoling1} Strong enough 
distortions can suppress the primary nonlinear phenomena and cause unwanted parasitic nonlinear
processes. The well-spread linear poling in commercially available samples of LiNbO$_3$
and LiTaO$_3$ crystals~\cite{LinearPoling,Review16}, is not appropriate for the comb generation. 
Among the most dangerous perturbation of the $2\pi/N$ periodicity is off-centering of the radial 
domain patterns. It can lead to noticeable side Fourier harmonics neighboring to $d_N$. The
presence of domain walls in radially poled microresonators can cause additional
losses. Nevertheless, the $Q$ factors in mm-sized resonators reach values $\sim 10^8$~\cite{RadialPoling1,Q-factors}. 
Substantially smaller radially poled on-chip resonators possess so far quality factots $Q \lesssim 10^6$~\cite{OnChip1,OnChip2}. Manufacturing of $\chi^{(2)}$ microresonators with high-quality radial 
poling and minimized losses can be regarded as a significant step in realization of the $\chi^{(2)}$ 
frequency combs.

\section{Governing equations for the antiperiodic states}

As the starting point for our derivation procedure we use the following generic nonlinear equations 
for complex modal amplitudes $F_{m_1}$ and $S_{m_2}$ relevant to FH and SH frequency ranges, respectively, 
and pumping into a SH mode with azimuth number $m^0_2$:
\begin{eqnarray}\label{Initial}
&{\rm i}& \hspace*{-2mm}\frac{dF_{m_1}}{dt} - (\omega_{m_1} - {\rm
i}\gamma_{m_1})F_{m_1} = 2\mu
\sum\limits_{m_2,m'_1} \; S_{m_2}F^*_{m'_1} \, \delta_{m_2 - m_1 - m'_1 - N} \nonumber  \\
& & \phantom{Q} \\
&{\rm i}& \hspace*{-2mm} \frac{dS_{m_2}}{dt} - (\omega_{m_2} - {\rm
i}\gamma_{m_2})S_{m_2} = \mu \sum\limits_{m_1,m'_1} \; F_{m_1}F_{m'_1} \, \delta_{m_2 -
m_1 - m'_1 - N} + {\rm i}h\delta_{m_2 - m^0_2}\,e^{-{\rm i}\omega_pt} \; . \nonumber
\end{eqnarray}
Here $\gamma_{m_{1,2}}$ are the modal decay parameters, such that the quality factors
$Q_{m_{1,2}} = \omega_{m_{1,2}}/2\gamma_{m_{1,2}} \gg 1$, $\mu$ is the coupling
coefficient incorporating the relevant susceptibility coefficient and modal overlaps~\cite{IlchenkoReviewI,Generic},
$\omega_p$ is the pump frequency, $h$ is a variable parameter characterizing pump
strength, and the asterisk indicates complex conjugation. The Kronecker symbols express
the momentum conservation law in the presence of radial poling. The amplitudes are
normalized such that $\omega_{m_1}|F_{m_1}|^2$ and $\omega_{m_2}|S_{m_2}|^2$ are the
modal energies. Without loss of generality, $\mu$ and $h$ can be treated as real
positive quantities. The FH and SH modes can be polarizationally the same or different.
The case $N = 0$ corresponds to phase matching of differently polarized modes without
radial poling (the natural phase matching~\cite{NaturalPM}). Transfer to the case of FH 
pumping is evident. The presence of a factor of $2$ in Eqs.~(\ref{Initial}) reflects the
Hamiltonian nature of classical $\chi^{(2)}$ interactions in the absence of modal 
losses~\cite{Hamiltonian}. In essence, set~(\ref{Initial}) accounts in a general manner 
for the effects of modal dispersion, decay, dissipativeless $\chi^{(2)}$ coupling, and 
frequency selective pumping. 

Now we assume that $m^0_2 - N = 2m^0_1 + 1$ is an odd number. Also we assume that the
pump frequency $\omega_p$ is very close to the SH frequency $\omega_{m^0_2}$ and
$\omega_{m^0_2}$ is very close to $\omega_{m_1^0} + \omega_{m^0_1 + 1}$. More
specifically, we suppose that the pump detuning $\Delta_p = \omega_{m^0_2} - \omega_p$
and the detuning $\Delta_q = \omega_{m^0_1} + \omega_{m^0_1 + 1} - \omega_{m^0_2}$
relevant to a slightly imperfect quasi-phase matching are much smaller in the absolute
values as compared to the intermodal distances $v_{1,2}/R$. At the same time, these
detunings can be comparable with or even larger than the FH and SH line widths 
$2\gamma_{m_{1,2}}$. The detunings $\Delta_{p}$ and $\Delta_q$ can be controlled 
separately in experiment.

Further simplifications imply relative narrowness of the FH and SH spectra. Let the
modal numbers $m_1$ and $m_2$ in Eqs.~(\ref{Initial}) be close to $m^0_1$ and $m^0_2$,
respectively, such that the deviations $\delta m_1 = m_1 - m^0_1$ and $\delta m_2 = m_2
- m^0_2$ are substantially smaller than $m^0_1$ and $m^0_2$. Then we can employ the Tailor 
expansions for the modal frequencies
\begin{equation}\label{Expansions}
\omega_{m_1} = \omega_{m^0_1} + \frac{v_1}{R} \,\delta m_1 + \frac{v'_1}{2R^2}\, \delta
m_1^{\; 2} , \qquad \omega_{m_2} = \omega_{m^0_2} + \frac{v_2}{R}\, \delta m_2 +
\frac{v'_2}{2R^2} \,\delta m_2^{\; 2} \;,
\end{equation}
where $v_{1,2}$ and $v'_{1,2}$ are the group velocities and the group velocity
dispersions for modes $m^0_{1,2}$, as defined and quantified in the main text. 
Analogous discrete expansions are wide-spread in the comb literature, see, e.g.,~\cite{KippNP14,KippScience18,Wabnitz18,Skryabin19}. Similar expansions can, 
generally speaking, be employed for the modal decay parameters $\gamma_{m_{1,2}}$.
Since the $m$-dependences of these parameters are typically less important than 
the effects of frequency dispersion, we set for simplicity $\gamma_{m_{1,2}} = 
\gamma_{1,2} = const$.

Using the above assumptions, it is not difficult to obtain evolutional partial
differential equations for true FH and SH envelopes $\mathcal{F}(\varphi,t)$ and
$\mathcal{S}(\varphi,t)$, that are given by the Fourier expansions
\begin{equation}
\mathcal{F} =  \sum\limits_{\delta m_1} F_{m_1}\, e^{{\rm i}\delta m_1 \varphi + {\rm
i}\omega_pt/2} = \sum\limits_{\delta m_1} \tilde{F}_{\delta m_1}\, e^{{\rm i}\delta m_1
\varphi}\;,   \qquad \mathcal{S} = \sum\limits_{\delta m_2} S_{m_2} \, e^{{\rm i}\delta
m_2 \varphi + {\rm i}\omega_pt} = \sum\limits_{\delta m_2} \tilde{S}_{\delta m_2} \;
e^{{\rm i}\delta m_2 \varphi}
\end{equation}
and $2\pi$-periodic in $\varphi$. The slow modal amplitudes $\tilde{F}_{\delta m_1}$ and
$\tilde{S}_{\delta m_2}$ obey Eqs.~(\ref{Initial}) if we replace in these equations
$\omega_{m_1} \to \omega_{m_1} - \omega_p/2$, $\omega_{m_2} \to \omega_{m_2} -
\omega_p$, $\delta_{m_2 - m_1 - m'_1 - N} \to \delta_{\delta m_2 - \delta m_1 - \delta
m'_1 + 1}$ and omit the factor $\exp(-{\rm i}\omega_pt)$. Multiplying the equations for
$\tilde{F}_{\delta m_1}$ and $\tilde{S}_{\delta m_2}$ by $\exp({\rm i} \delta
m_1\varphi)$ and $\exp({\rm i}\delta m_2\varphi)$, respectively, taking the sums in
$\delta m_1$ and $\delta m_2$, and employing Eqs.~(\ref{Expansions}), we obtain:
\begin{eqnarray}\label{True}
\left({\rm i}\partial_t + {\rm i}\frac{v_1}{R}\,\partial_{\varphi} + \frac{v'_1}{2R^2}
\,\partial^2_{\varphi} + \frac{\omega_p}{2} - \omega_{m^0_1} + {\rm i}\gamma_1
\right)\mathcal{F} &=&
2\mu \; \mathcal{S}\mathcal{F}^*\,\exp({\rm i} \varphi) \nonumber  \\
& & \phantom{Q} \\
\left({\rm i}\partial_t + {\rm i}\frac{v_2}{R}\,\partial_{\varphi} + \frac{v'_1}{2R^2}\,
\partial^2_{\varphi} + \omega_p - \omega_{m^0_2} + {\rm i}\gamma_2     \right)\mathcal{S} &=&
\mu \; \mathcal{F}^2 \,\exp(-{\rm i} \varphi) + {\rm i} h \;. \nonumber
\end{eqnarray}
Explicit dependences of the right-hand sides on the azimuth angle $\varphi$ make
Eqs.~(\ref{True}) not autonomous and, thus, inconvenient. This feature is a direct
consequence of the phase-matching conditions used. However, we can make
Eqs.~(\ref{True}) autonomous by substituting $\mathcal{F}(\varphi,t) =
F(\varphi,t)\exp(i\varphi/2)$. Setting for uniformity $\mathcal{S}(\varphi,t) =
S(\varphi,t)$, we arrive at the following autonomous generic set of nonlinear equations~\cite{We19}:
\begin{eqnarray}\label{FinalSet}
\left({\rm i}\partial_t + {\rm i}\frac{v_1}{R}\,\partial_{\varphi} + \frac{v'_1}{2R^2}
\,\partial^2_{\varphi} - \Delta_1 + {\rm i}\gamma_1 \right) F &=& 2\mu \; S F^*  \nonumber \\
& & \phantom{Q} \\
\left({\rm i}\partial_t + {\rm i}\frac{v_2}{R}\,\partial_{\varphi} + \frac{v'_1}{2R^2}\,
\partial^2_{\varphi} - \Delta_2 + {\rm i}\gamma_2 \right)S &=&
\mu \; F^2 + {\rm i} h  \nonumber
\end{eqnarray}
for $2\pi$-periodic SH amplitude $S$ and $2\pi$-antiperiodic FH amplitude $F$. In this
set, the detunings $\Delta_{1,2}$ are given by the expressions $\Delta_1 = (\Delta_p +
\Delta_q - v'_1/4R^2)/2$ and $\Delta_2 = \Delta_p$, where, as earlier, $\Delta_p =
\omega_{m^0_2} - \omega_p$ and $\Delta_q = \omega_{m^0_1} + \omega_{m^0_1 + 1} -
\omega_{m^0_2}$ are the pump and quasi-phase-match detunings. By setting $\Delta_p = 0$
and varying $\Delta_q$, we can always achieve the case $\Delta_{1,2} = 0$ relevant to
our main results. Note that setting $\Delta_p = \Delta_q = 0$ does not turn the detuning
$\Delta_1$ exactly to zero. However, the remnant detuning $-v'_1/8R^2$ is typically
negligible as compared to the FH line width $2\gamma_1$.

Setting $\Delta_{1,2} = 0$ and $\gamma_1 = \gamma_2 = \gamma$ in Eqs.~(\ref{FinalSet}), transferring 
to the coordinate frame moving with velocity $v_1$ and to the normalized FH and SH amplitudes $f = (\mu/\gamma) (2i)^{1/2}F$ and $s = (2\mu/\gamma) S$, and using the expansions
\begin{equation}\label{fs-expansions}
f(\varphi,t) = \sum\limits_{j_1} f_{j_1}(t)\,e^{{\rm i}j_1\varphi}, \qquad s(\varphi,t) = \sum\limits_{j_2} s_{j_2}(t)\,e^{{\rm i}j_2\varphi}
\end{equation}
with $j_1 = \pm 1/2,\pm 3/2, \ldots$ and $j_2 = 0, \pm 1,\pm 2, \ldots$, we obtain a set of coupled 
ordinary differential equations for harmonics $f_{j_1}$ and $s_{j_2}$:
\begin{eqnarray}\label{CoupledHarmonics}
\left(\frac{1}{\gamma}\frac{d}{dt} + {\rm i}\beta_1j_1^2 + 1 \right)f_{j_1} &=& \sum\limits_{j'_1} f^*_{j'_1}s_{j_1 + j'_1} \\
\left(\frac{1}{\gamma}\frac{d}{dt} - {\rm i}\alpha j_2 + {\rm i}\beta_2j_2^2 + 1\right)s_{j_2} &=& 
\eta \, \delta_{j_2} - \sum\limits_{j'_1} f_{j'_1}f_{j_2 - j'_1} \;. \nonumber
\end{eqnarray}
Here $\alpha = v_{12}/\gamma R$, $\beta_{1,2} = v'_{1,2}/2\gamma R^2$, and $\eta = 2\mu h/\gamma^2$ are 
dimensionless parameters. The product $\gamma t$ can be treated as dimensionless time. Set~(\ref{CoupledHarmonics}) has to be solved with certain initial conditions for $f_{j_1}$ and $s_{j_2}$. 
Achievement of a steady state corresponding to movement of the FH and SH envelopes $f(\varphi,t)$ and 
$s(\varphi,t)$ with a common velocity $v_0 \neq v_1$ means that the steady-state amplitudes oscillate harmonically in time as $\exp(-{\rm i}jv_{01}t/R)$.

\section{Establishment of steady states: the quasi-adiabatic procedure}

Here we exhibit representative data of numerical calculations relevant to transient processes and achievement 
of antiperiodic steady states for the FH and SF envelopes. Our quasi-adiabatic calculation procedure 
was organized as follows: The ordinary differential equations for $512$ coupled FH and SH harmonics 
were solved in the time domain for different values of the walk-off parameter $\alpha$. The pump strength parameter $\eta$ was increased stepwise from $\eta_1 = 1.01$ to $30$. For each discrete value $\eta_s$ 
with $s = 1,2,\ldots$, we calculated velocity $v_0(\eta_s)$ and the discrepancy parameter $\varepsilon(t)$, 
as defined in the main text, and made sure that it is saturated by $t = t_s$ on an extremely low level ($\varepsilon \sim 10^{-14} - 10^{-15}$). The values of the complex amplitudes $F_{j_1}$ and $S_{j_2}$ achieved by time $t_s$ were used as initial conditions for a new calculation run 
with $\eta_{s + 1} > \eta_s$. This quasi-adiabatic iteration procedure was repeated untill the maximum value $\eta = 30$ was achieved. For the first calculation step, relevant to $\eta_1 = 1.01$, we started from a weak random noise, $|f_{j_1}|^2,|s_{j_2}|^2 \lesssim 10^{-10}$ for $j_2 \neq 0$ and $s_0 = \eta_1$. After reaching 
$\eta_{\max} = 30$, we repeated the quasi-adiabatic procedure by decteasing $\eta$ and made sure that the same steady states were passed in reverse order. 

\begin{figure}[h]
\centering
\includegraphics[width=12.2cm]{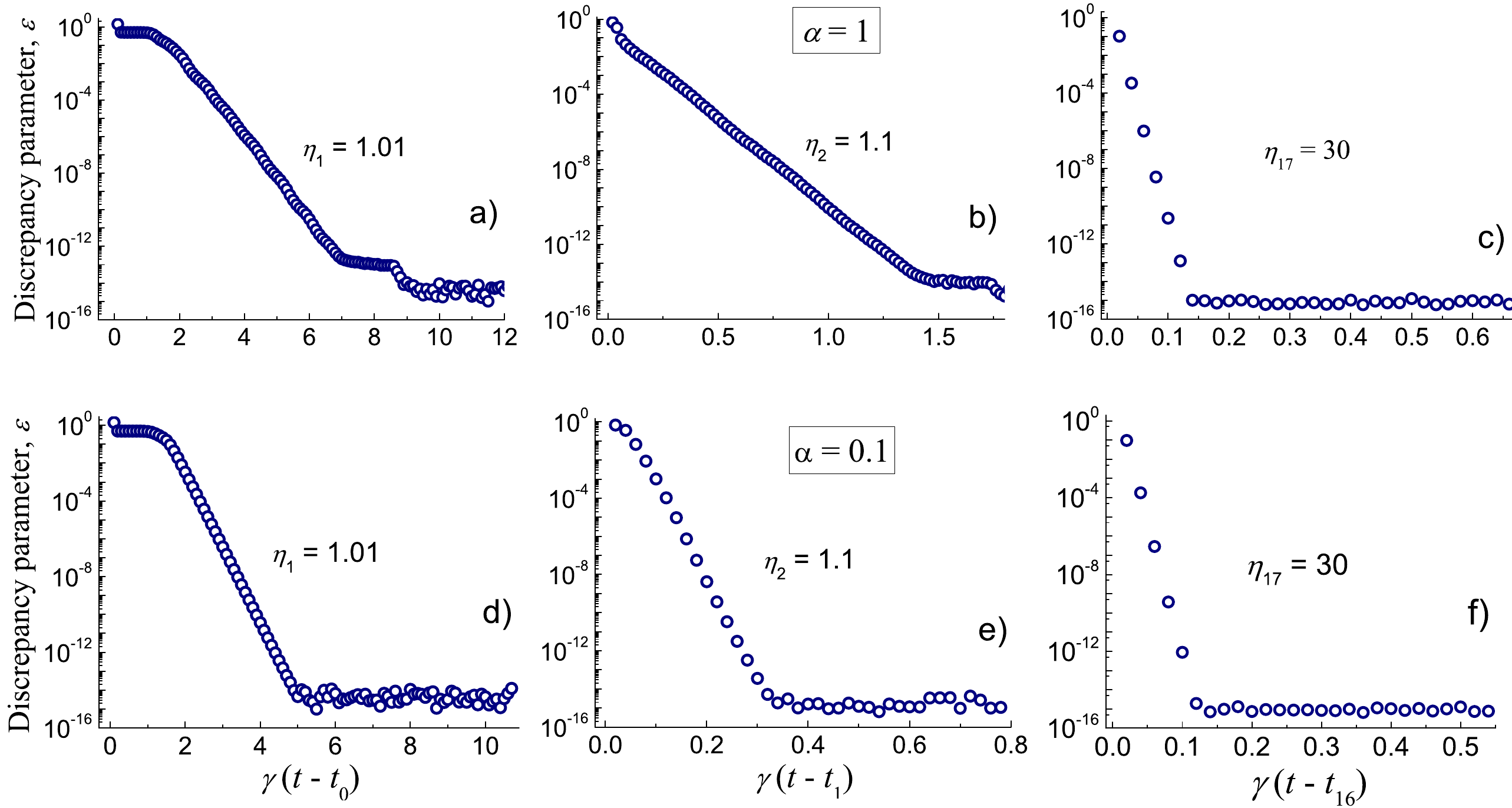}
\caption{Six representative evolution runs for the discrepancy parameter $\varepsilon(t,\tau)$ at 
$\gamma \tau = 1$; the time step is $0.02\gamma^{-1}$. The first and second rows correspond to $\alpha = 1$ and $0.1$, respectively. The evolutions relevant to a) and d) start at $t_0$ from noise and end up at $t_1$.  512 harmonics are taken into account and the model paramers used in the simulations are: $\gamma = 10^7$~s$^{-1}$, $\beta_1 = 0.02$, and $\beta_2 = -0.01$.  } \label{Trans}
\end{figure}

Figure~\ref{Trans} shows representative examples of temporal evolution of the FH discrepancy parameter $\varepsilon(t,\tau)$ for the time shift $\tau = 1/\gamma$. The first and second rows correspond to
the walk-off parameter $\alpha = 1$ and $0.1$. 
The chosen 17 values $\eta_s$ are $1.01, 1.1, 1.3, 2., 2.5, 3, 3.5, 4, 5, 7, 10, 12, 15, 18, 20, 25$, and $30$. Subfigures a), b), c) and also d), e), f) exhibit evolution of $\varepsilon$ for $\eta_1$, $\eta_2$, and $\eta_{17}$. The most general feature this evolution is always the same: It is an almost exponential decrease of $\varepsilon$ from initial values of the order of $1$ to extremely low values $10^{-14} - 10^{-15}$ caused by the numerical noise. Quantitative details of this evalution are also important. The longest evolution, subfigures a) and d), starts from modal noise. It takes  $\sim (5-7)\gamma^{-1}$ for the harmonics to grow exponentially and reach the saturation. The corresponding evolution time can be identified with the rise time of near-threshold optical parametric oscillation in microresonators. The further evolution runs, which start from regular distributions of the FH and SH harmonics achieved during the previous runs, occur substantially faster. Our choice of $17$ steps of evolution to reach the soliton states with $\eta = 30$ is largerly arbitrary. In many cases, especially for $\alpha \gtrsim 1$, the same final state can be achieved even in one step.  Variations of the time shift $\tau$ give no effect on the data of Fig.~\ref{Trans}. The temporal evolution of the SH discrepancy parameter behaves in a similar manner.

Anyhow, the numerical simulations evidence that our coherent antiperiodic soliton states, represented by 
Figs.~4-6 of the main text, are very robust. Not too abrupt turning the pump power on, with the pump rise time 
$t_p$ exceeding $10\gamma^{-1}$, ensures achievement of these comb states. This restriction, 
relevant actually to $t_p \gtrsim 1\,\mu$s, is very soft.